\documentclass[pre,aps,groupaddress,showpacs,prd,twocolumn]{revtex4}
\usepackage{epsfig,amsmath,graphicx,amssymb,overpic}

\def\be{\begin{equation}}
\def\ee{\end{equation}}
\def\bee{\begin{eqnarray}}
\def\ene{\end{eqnarray}}
\def\bes{\begin{subequations}}
\def\ees{\end{subequations}}

\begin{document}
%%%%%%%%%%%%%%%%%%%%%%%%%%%%%%%%%%%%%%%%%%%%%%%%%%%%%%%%%%%%%%%%%%%%%%%%%%%%%%%%%
\title{Optical rogue waves in the generalized inhomogeneous higher-order \\ nonlinear Schr\"odinger equation with modulating coefficients}
\author{Zhenya Yan$^{1}$}
\email{zyyan\_math@yahoo.com}
\author{Chaoqing Dai$^{2}$}

\affiliation{ \vspace{0.05in} \rm $^{1}$Key Laboratory of Mathematics Mechanization, Institute of
Systems Science, AMSS,  Chinese Academy of Sciences, Beijing 100190, China. \\
$^{2}$School of Sciences, Zhejiang Agriculture and Forestry University, Lin¡¯an, Zhejiang CN-311300, China}

%\date{\today}

%%%%%%%%%%%%%%%%%%%%%%%%%%%%%%%%%%%
 \vspace{0.1in}
\begin{abstract}

\noindent  {\bf Abstract}  \vspace{0.02in}\\
The higher-order dispersive and nonlinear effects (alias {\it the perturbation terms}) like the third-order dispersion, the self-steepening, and the self-frequency shift play important roles in the study of the ultra-short optical pulse propagation. We consider optical rogue wave solutions and interactions for the generalized higher-order nonlinear Schr\"odinger (NLS) equation with space- and time-modulated parameters. A proper transformation is presented to reduce the generalized higher-order NLS equation to the integrable Hirota equation with constant coefficients. This transformation allows us to relate certain class of exact solutions of the generalized higher-order NLS equation to the variety of solutions of the integrable Hirota equation.
In particular, we illustrate the approach in terms of two lowest-order rational solutions of the Hirota equation as seeding functions to generate rogue wave solutions localized in time that have complicated evolution in space with or without the differential gain or loss term. We simply analyze the physical mechanisms of the obtained optical rogue waves on the basis of these constraints. Finally, The stability of the obtained rogue-wave
solutions is addressed numerically. The obtained rogue wave solutions may raise the possibility of relative experiments and potential applications in nonlinear optics and other fields of nonlinear science as Bose-Einstein condensates and ocean.\vspace{0.02in}

\noindent {\bf Keywords:} Higher-order effects, the ultra-short optical pulse, generalized higher-order NLS equation, optical rogue waves, Hirota equation

\end{abstract}
%\pacs{05.45.Yv, 03.75.Lm, 42.65.Tg, 42.50.Gy}

\maketitle

%%%%%%%%%%%%%%%%%%%%%%%%%%%%%%%%%%%%%%%%%%%%%%%%%%%%%%%%%%%%%%%%%

%%%%%%%%%%%%%%%%%%%%
%%%%%%%%%%%%%%%%%%%%

\section{Introduction}

The laser beam intensity-dependent refractive index of the optical Kerr effect $n(I)=n_0+n_2 I$  can generate
the nonlinear Schr\"odinger (NLS) equation
\bee\label{nls}
 i\frac{\partial \psi}{\partial z}+\frac{\beta}{2}\frac{\partial^2 \psi}{\partial t^2}+g|\psi|^2\psi=0
\ene
describing the propagation of light within an optical fiber or a
uniform nonlinear medium~\cite{op,op2,op3,op4}, where $n_0$ denotes the linear refractive index, $n_2$ stands for a Kerr-type
nonlinearity of the waveguide amplifier, $I$ is the laser beam intensity, $\beta$ denotes the group velocity dispersion (GVD), and
 $g$ is the Kerr nonlinearity (or self-phase modulation). Eq.~(\ref{nls}) is also called the Gross-Pitaevskii equation without the external potential in Bose-Einstein condensates~\cite{bec,bec2,bec3,bec4} if one replaces $z$ by $t$ and $t$ by $x$. Eq.~(\ref{nls}) has been shown to be completely integrable~\cite{soliton} and to admit
optical solitons by balancing GVD $\beta$ and Kerr nonlinearity $g$ (the self-focusing interaction and defocusing interaction corresponding to bright and dark solitons, respectively)~\cite{op,soliton,ch}. For the varying second-order GVD $\beta\to\beta(z)$ and the self-phase modulation (SPM) $g\to g(z)$, Eq.~(\ref{nls}) describes the propagation of optical pulses in a self-similar manner in a
nonlinear waveguide~\cite{sl, sl2, sl3}, and is shown to have the novel (self-similar) optical solitons~\cite{nlsvc,nlsvc1,nlsvc2} including some free functions of space $z$, which generates complicated wave propagations. The nonlinear self-similar wave implies that its profile remains unchanged
and its amplitude and width simply scale with time or propagation distance (see, e.g., \cite{nlsvc2, pg1,self1,self2,self3,self4}). If one exchanges the variables
$z$ and $t$, then this kind of novel solitons is call nonautonomous solitons (see, e.g., \cite{p3}).
Other generalized models of Eq.~(\ref{nls}) with varying parameters have been studied such as the varying external potential~\cite{p1,p2,p3,pg1,pg2}, the varying gain or loss term~\cite{pg1,pg2,3d1}, and three-dimensional cases~\cite{3d1,3d2,3d3}, and shown to admit exact solutions and complicated wave propagations.

When the optical pulses become shorter (e.g., 100 fs~\cite{op}), the higher-order dispersive and nonlinear effects like the
third-order dispersion (TOD), the self-steepening (SS), and the self-frequency shift (SFS) (alias {\it the perturbation terms}) arising from the stimulated Raman scattering become significant in the study of the ultra-short optical pulse propagation (see, e.g.,~\cite{op}). To the above-mentioned aim Kodama and Hasegawa~\cite{nls3a,nls3b} presented the higher-order NLS equation
\bee\label{nls3}
\begin{array}{l}
\displaystyle i\frac{\partial \psi}{\partial z}+\frac{\beta}{2}\frac{\partial^2 \psi}{\partial t^2}+g|\psi|^2\psi\qquad\quad  \vspace{0.08in}\cr
\displaystyle \qquad =i\epsilon\!\left[a_1\frac{\partial^3 \psi}{\partial t^3}\!+\!a_2\frac{\partial (|\psi|^2\psi)}{\partial t}\!+\!a_3
 \psi\frac{\partial |\psi|^2}{\partial t}\right]+i\Gamma\psi,\quad
 \end{array}
\ene
reduced from the Maxwell equation $\nabla\times \nabla\times {\bf E}=-\frac{1}{c^2}\frac{\partial^2 {\bf D}}{\partial t^2}$ with $c$ being the speed of light and ${\bf D}=\epsilon \ast {\bf E}$~\cite{nls3b}, where ${\bf E}$ is the electric field intensity, $\epsilon$ is a real-valued parameter,  these real-valued parameters $\beta,\, g,\,a_1,\,a_2,\,a_3,$ and $\Gamma$ are related to GVD, SPM, SS, SFS, and the gain or loss terms, respectively.  Eq.~(\ref{nls3}) contains many types of intergrable models such as the Hirota equation~\cite{nlsh}, Sasa-Satsuma equation~\cite{nlss}, KN-type derivative NLS equation~\cite{nlsk}, and CLL-type derivative NLS equation~\cite{nlsc}.
 Eq.~(\ref{nls3}) has been found to admit many types of solutions (see, e.g.,~\cite{hnlsc,li, gi, vy}). For the ultra-short pulse in a self-similar matter, Eq.~(\ref{nls3}) with varying coefficients has been shown to support exact solutions for some constraints of coefficients (see, e.g.,~\cite{hc1,hc2,hc3,hc4,hc5}). The experimental observations are recently performed for optical solitons and pulse compression in
 $\sim\!1$-mm-long photonic crystal waveguides~\cite{hnlse} based on the higher-order NLS model~\cite{hnlst}.

Recently, the NLS equation and its extensions have been paid much attention due to the interesting
phenomenon of rogue waves, which are localized both in space and time and depict a unique event that appears
from nowhere and disappears without a trace~\cite{na}. Rogue waves are also known as {\it freak waves}, {\it monster
waves}, {\it killer waves}, {\it gaint waves} or {\it extreme waves}.  In particular, optical rogue waves have been experimentally verified in nonlinear optics which play an important role in the supercontinuumg generation~\cite{orw,orw2} on the basis of the generalized NLS equation with
$N=3,4,5,6$
 \bee\nonumber
\begin{array}{l}
\displaystyle i\frac{\partial A}{\partial z}+\sum_{n=2}^N\frac{i^n\beta_n}{n!}\frac{\partial^n A}{\partial t^n}  \vspace{0.08in}\cr
\displaystyle \qquad =-g\!\left[|A|^2A+\frac{i}{\omega_0}\frac{\partial (|\psi|^2\psi)}{\partial t}-T_RA\frac{\partial |\psi|^2}{\partial t}\right],\quad
 \end{array}
\ene
where $A(z,t)$ denotes the field, $\beta_n$ are real values characterizing the fibre dispersion, $g$ is the nonlinear
coefficient of the fibre, $\omega_0$ is the central carrier frequency of the field, and $T_R$
is a parameter that characterizes the delayed nonlinear response of silica fibre.

 Moreover, optical rogue waves were theoretically studied in telecommunication data streams~\cite{orwt}. Except for the optical rogue waves, rogue wave
phenomena also appear in ocean~\cite{srw,srw2,srw3,srw4}, Bose-Einstein condensates~\cite{mrw,mrw2}, and even finance~\cite{frw,frw2}. Rogue wave solutions have analytically been found for many types of nonlinear physical models such as  nonlinear models with constant coefficients (e.g., the NLS equation~\cite{PS,nlsr}, higher-order NLS equation~\cite{hnls,lilu}, Hirota equation~\cite{hr}), the discrete models (e.g., discrete Ablowitz-Ladik and Hirota equations~\cite{drw,drw2}, the generalized Ablowitz-Ladik-Hirota lattice with varying coefficients~\cite{drw3}), nonlinear models with varying coefficients (e.g., the NLS equation with varying coefficients~\cite{pg2}, three-dimensional NLS equation with varying coefficients~\cite{mrw2}), and the coupled NLS equations~\cite{frw2,crw,crw2}. Recently, the controllable rogue waves of Eq.~(\ref{nls3}) with all coefficients depending on only space $z$ have been considered~\cite{hrw}, but
one more condition was missed to support the obtained results, that is to say, the sum of the parameters related to the self-steepening and the self-frequency shift should be zero ($a_2+a_3=0$).

Eq.~(\ref{nls3}) with coefficients depending on only space $z$ was studied before (see, e.g.,~\cite{hc1,hc2,hc3,hc4,hc5,hrw}. A natural problem is what waves (space, time)-modulated GVD and SPM terms can make Eq.~(\ref{nls3}) excite. Moreover, could other terms (e.g., the additional external potential~\cite{p1,p2,p3,pg1,pg2}, the differential gain or loss parameter~\cite{ghnls}, and the group velocity~\cite{op,chow,hnlst}) excite Eq.~(\ref{nls3}) to generate new phenomena ? Based on these motivations related to some physical phenomena, in this paper we study the generalized model of Eq.~(\ref{nls3}) with (space, time)-modulated GVD, SPM, and gain or loss terms, space-modulated TOD, SS and SFS, and more terms like the (space, time)-modulated external potential, linear group velocity, and differential gain or loss term~\cite{op,ghnls}
\bee \label{chnls}
\begin{array}{l}
\displaystyle i\frac{\partial \psi}{\partial z}\!=\!\beta(z,t)\frac{\partial^2 \psi}{\partial t^2}\!+\!\left[V(z,t)\!+\!i\gamma(z,t)\right]\psi
\!+\!g(z,t)|\psi|^2\psi  \vspace{0.08in}\cr
\displaystyle \quad\quad\,\,
 + i\!\!\left[\alpha_1(z)\frac{\partial^3 \psi}{\partial t^3}\!+\!\alpha_2(z)\frac{\partial (|\psi|^2\psi)}{\partial t}\!+\!\alpha_3(z)\psi\frac{\partial |\psi|^2}{\partial t}\right] \quad \vspace{0.08in}\cr
\displaystyle \quad\quad\,\, +\left[\mu(z)+i\sigma(z,t)\right] \frac{\partial \psi}{\partial t},
 \end{array}\ene
where  $z$ is the normalized propagation distance along the optical fibre, $t$ is the retarded time, $\psi\equiv \psi(z,t)$ denotes  the slowly varying envelope amplitude of the electric field measured in units of square root of Watts at position $z$ in the optical fiber and at time $t$,
$\beta(z,t),\, g(z,t),\, \alpha_1(z), \alpha_2(z)$, and $\alpha_3(z)$ are all real-valued functions of displayed variables and stand for GVD, SPM, TOD, SS, and SFS  arising from
stimulated Raman scattering, respectively~\cite{op}, $V(z,t)$ and $\gamma(z,t)$ are the external potential and gain or loss distribution, respectively, $\mu(z)$ denotes the differential gain or loss parameter~\cite{ghnls}, and $\sigma(z,t)$ is related inversely to the group
velocity of the modes (a ¡®walk-off¡¯ effect)~\cite{op,chow,hnlst}. Eq.~(\ref{chnls}) is associated with a variational principle  $\delta \mathcal{L}/\delta\psi^{*}=0$  with the Lagrangian density
\bee\begin{array}{l}
\mathcal{L}=i(\psi\psi_z^{*}-\psi_z\psi^{*})-2\beta(z,t)|\psi_t|^2+g(z,t)|\psi|^4 \vspace{0.1in} \cr \qquad
                     +2[V(z,t)+i\gamma(z,t)]|\psi|^2+\alpha_1(z)(\psi_t\psi_{tt}^{*}-\psi_t^{*}\psi_{tt}) \quad
                     \vspace{0.1in} \cr \qquad
                     +[\mu(z)+i\sigma(z,t)](\psi_t\psi^{*}-\psi\psi^{*}_t)
\vspace{0.1in} \cr \qquad
+[\alpha_2(z)(|\psi|)_t+\alpha_3(z)\psi\psi^{*}_t]|\psi|^2,
\end{array}
\ene
where $\psi^*$ stands for the complex conjugation of the electric field $\psi$, and the subscript denotes the partial derivative with respect to the variables ($z,t)$. Eq.~(\ref{chnls}) contains many types of nonlinear models
such as the NLS equation with varying coefficients~\cite{nlsvc, nlsvc1, nlsvc2}, the derivative NLS equation with varying coefficients~\cite{nlsk,nlsc}, the Hirota equation with varying coefficients, the Sasa-Satsuma equation with varying coefficients~\cite{hc1,hc2,hc3,hc4,hc5}), and the higher-order NLS equation without three photon
nonlinear absorption~\cite{hnlst}.

 The rest of this paper is organized as follows. In Sec. II, we present a proper transformation reducing Eq.~(\ref{chnls}) to the Hirota equation and determine the similarity variables and constraints satisfied by the external potential, GVD, SPM, TOD, SS, SFS, and gain or loss terms in Eq.~(\ref{chnls}). Sec. III mainly focuses on two types of time-localized rogue wave solutions for some chosen parameters. Moreover, we analyze the obtained first-order optical rogue wave solutions of Eq.~(\ref{chnls}) by using numerical simulations. Finally, these results and discussions are summarized in the conclusion.

\section{Symmetry reductions and solutions}

\subsection{Symmetry reductions}

In general, Eq.~(\ref{chnls}) is not integrable since these varying coefficients strongly affect on the wave propagation of optical pulses in a self-similar manner. In order to study exact analytical solutions of Eq.~(\ref{chnls}), we need to seek for its some integrability conditions using some methods like the symmetric reduction approach, Painlev\'e analysis (see, e.g., \cite{soliton,lie,lie2} and references therein). Here we consider the symmetry reductions of Eq.~(\ref{chnls}). Eq.~(\ref{chnls}) has many different types of symmetry reductions by employing the Lie group transformation method (see, e.g.,~\cite{lie,lie2}) to Eq.~(\ref{chnls}) which means that these similarity reductions can be obtained using the third order propagation ${\rm Pr}^{(3)}(\chi)$ of the vector field (also called the infinitesimal generator)
\bee
\chi=T\frac{\partial}{\partial t}+Z\frac{\partial}{\partial z}+\theta_R\frac{\partial}{\partial \psi_R}+
\theta_I\frac{\partial}{\partial \psi_I}
\ene
acting on Eq.~(\ref{chnls}), i.e., ${\rm Pr}^{(3)}(\chi)F\big|_{\{F=0\}}=0$,
 where $\psi(z,t)=\psi_R(z,t)+i\psi_I(z,t)$ with $\psi_R,\,\psi_I\in \mathbb{R}$, these variables $T,\, Z,\, \theta_R,$ and $\theta_I$ are all unknown functions of $z, t, \psi_I, \psi_R$ to be determined, and $F=\{-i\partial_z+\beta\partial_t^2+(V+i\gamma)+g|\psi|^2+i[
\alpha_1\partial_t^3+\alpha_3\partial_t(|\psi|^2)]+(\mu+i\sigma)\partial_t\}\psi+\alpha_2\partial_t(\psi|\psi|^2)\big|_{\psi=\psi_R+i\psi_I}
 $, ${\rm Pr}^{(3)}(\chi)=\chi+\theta_R^z\frac{\partial}{\partial \psi_{R,z}}+\theta_I^z\frac{\partial}{\partial \psi_{I,z}}+\theta_R^t\frac{\partial}{\partial \psi_{R,t}}+\theta_I^t\frac{\partial}{\partial \psi_{I,t}}
 +\theta_R^{tt}\frac{\partial}{\partial \psi_{R,tt}}+\theta_I^{tt}\frac{\partial}{\partial \psi_{I,tt}}
 +\theta_R^{ttt}\frac{\partial}{\partial \psi_{R,ttt}}+\theta_I^{ttt}\frac{\partial}{\partial \psi_{I,ttt}}$. The similarity variables and transformations can be founded by solving the characteristic
equation $\frac{dt}{T}=\frac{dz}{Z}=\frac{d\psi_R}{\theta_R}=\frac{d\psi_I}{\theta_I}$~\cite{lie,lie2}.

Here, to study multi-soliton solutions of Eq.~(\ref{chnls}) including rogue wave solutions, we need to reduce it to some integrable (or easily solved) differential equations (e.g., Hirota equation~\cite{nlsh}, Sasa-Satsuma equation~\cite{nlss}, which possess multi-soliton solutions or even rogue waves). In the following we consider the symmetry (reduction) transformation~\cite{pg2,mrw2}
\bee
 \psi(z,t)=\rho(z)e^{i\varphi(z,t)}\Psi[\tau(z,t), \eta(z)]
   \label{trans}
 \ene
connecting solutions of Eq.~(\ref{chnls}) with those of the following
Hirota equation with constant coefficients~\cite{nlsh}
\bee\label{3nls}
 i\frac{\partial\Psi}{\partial \eta}\!=\!-\frac{\partial^2\Psi}{\partial \tau^2}\!+\!G|\Psi|^2\Psi\!+\!2\sqrt{2}i\nu\!\left(\frac{\partial^3\Psi}{\partial \tau^3}
 \!+\!3|\Psi|^2\frac{\partial\Psi}{\partial \tau}\!\right)\!, \quad\,\,
\ene
where the physical fields $\Psi(\eta, \tau)$ is the function
of the variables $\tau(z,t)$ and $\eta(z)$, which are the new temporal and spatial coordinates, respectively,
and $G, \nu$ are both real-valued constants.  Since the
main goal of this paper focuses on rogue waves of Eq.~(\ref{chnls}), we choose $G<0$\, (e.g., $G=-1)$ in Eq.~(\ref{3nls})  which corresponds to
the attractive case (or self-focusing interaction in nonlinear optics~\cite{op} and
attractive interaction (negative scattering lengths) in the BEC theory~\cite{bec,bec2,bec3,bec4}).

Notice that the higher-order equation (\ref{3nls}) with $\nu\not=0$ differs from the NLS equation (\ref{nls}) which admits two basic transformations, i.e.,
the scaling and gauge transformations, leave itself invariant (see, e.g., \cite{ps}). We find Eq.~(\ref{3nls}) have the following proposition:

{\it Proposition-.} Eq.~(\ref{3nls})  possesses two modified basic transformations leaving itself invariant, i.e., the following scaling-parameter transformation
\bee\label{b1}
 \eta\to\alpha^2\eta, \quad \tau\to\alpha \tau,\quad \Psi\to \Psi/\alpha,\quad \nu\to \alpha\nu,
 \ene
 leaving itself invariant for any SPM parameter $G\not=0$, where $\alpha$ is a real-valued constants, and the `gauge' transformation
\bee\label{b2}
 \eta\to k^3\eta, \quad \tau\to k\tau+\lambda\eta,\quad \Psi\to k^2\Psi e^{i(p\tau+q\eta)},
\ene
leaving Eq.~(\ref{3nls}) invariant only for the self-focusing SPM parameter $G=-1$, which just make Eq.~(\ref{3nls}) possess rogue wave solutions~\cite{hnlsrw}, where $k=1+6\sqrt{2}\nu p,\, \lambda=-kp(1+k), \, q=-p^2(2\sqrt{2}\nu p+1)$  with $p$ being a real-valued constant.

Based on the above-mentioned method, the substitution of Eq.~(\ref{trans}) into Eq.~(\ref{chnls}) with $\Psi(\eta,\tau)$ satisfying Eq.~(\ref{3nls})
yields the following system of partial differential equations
\bes \label{sysi} \bee
\label{sysi1} && \alpha_1\tau_t\tau_{tt}=0, \qquad  \alpha_2+\alpha_3=0, \\
\label{sysi2} && \beta\tau_{tt}-3\alpha_1(\tau_{tt}\varphi_t+\tau_{t}\varphi_{tt})+\mu\tau_t=0, \\
\label{sysi3} && \eta_z+\tau_t^2(\beta-3\alpha_1\varphi_t)=0, \\
\label{sysi4} && \sigma\tau_t+2\beta\tau_t\varphi_t+\alpha_1(\tau_{ttt}-3\tau_t\varphi_t^2)-\tau_z=0, \\
\label{sysi5} && \rho_z-\rho[\beta\varphi_{tt}-3\alpha_1\varphi_{t}\varphi_{tt}+\mu\varphi_t+\gamma]=0, \qquad\quad\\
\label{sysi6} && V+\varphi_z-\beta\varphi_t^2+\alpha_1(\varphi_t^3-\varphi_{ttt})-\sigma\varphi_t=0,\\
\label{sysi7} && \alpha_1\tau_t^3-2\sqrt{2}\nu\eta_z=0, \\
\label{sysi8} && \rho^2(g-\alpha_2\varphi_t)-G\eta_z=0,\\
\label{sysi9} && \alpha_2\rho^2\tau_t-6\sqrt{2}\nu\eta_z=0.
\ene\ees

Generally speaking, system (\ref{sysi}) may not be compatible
with each other, however, one can find some proper constraints
for these coefficients $\beta(z,t),\, g(z,t),\, V(z, t),\, \gamma(z,t),\, \alpha_j(t),\, \mu(z),$ and $\sigma(z,t)$
such that system (\ref{sysi}) is compatible. This requirement leads us to
the following procedure.

First of all, we solve Eqs.~(\ref{sysi1})-(\ref{sysi3}) to obtain the similarity variables $\tau(z, t),\, \eta(z)$, and
the phase $\varphi(z,t)$ subjec to the GVD parameter $\beta(z,t)$ the differential gain or loss term $\mu(z)$.

Secondly, it follows from Eqs.~(\ref{sysi4})-(\ref{sysi9}) that we can
determine the amplitudes $\rho(z)$, the external potential $V(z, t)$, SPM $g(z,t)$,\, GVD $\beta(z,t)$, TOD $\alpha_1(z)$,
$\sigma(z,t)$ SS $\alpha_2(z)$ and SFS $\alpha_3(z)$ in terms of the obtained variables $\eta(z),\, \tau(z,t)$, and $\varphi(z, t)$.

Finally, we may establish a `bridge' (also called Lie-B\"acklund transformation) between exact solutions of Eq.~(\ref{chnls})
and ones of completely integrable Hirota equation (\ref{3nls}). The latter admits an infinite number of solutions
thereby giving us an approach to find physically relevant
solutions of Eq.~(\ref{chnls}). Here we only consider optical rogue wave solutions of Eq.~(\ref{chnls})
based on the rogue waves of the Hirota equation~\cite{hnlsrw}.

\subsection{Determining similarity variables and controlled coefficients}

For the considered Eq.~(\ref{chnls}) in the presence of TOD, i.e., $\alpha_1(z)\not\equiv 0$, it follows from Eq.~(\ref{sysi1}) that the new temporal variable $\tau(z,t)$ should be of the form
 \bee\label{tau}
 \tau=\tau_1(z)t+\tau_0(z),
 \ene
 where $\tau_1(z)$ and $\tau_0(z)$ are functions of $z$. The substitution of Eq.~(\ref{tau}) into Eq.~(\ref{sysi2}) yields
 \bee
  (3\alpha_1\varphi_{tt}-\mu)\tau_t=0,
 \ene
 which leads to two cases since $\tau_t\not\equiv 0$ (i.e., $\tau_1(z)\not\equiv 0$ is required, ortherwise the field $\Psi(\eta,\tau)$ is only a function of $z$):

 \begin{itemize}

 \item[(I)] $\mu(z)\equiv 0,\, \varphi(z,t)=\varphi_1(z)t+\varphi_0(z)$;

  \item[(II)] $\mu(z)\not\equiv 0,\, \varphi(z,t)=\varphi_2(z)t^2+\varphi_1(z)t+\varphi_0(z)$ with $\mu(z)=6\alpha_1(z)\varphi_2(z)$.

\end{itemize}

 Based on the above-mentioned two cases, we have the following solutions of system (\ref{sysi}):

\vspace{0.1in}{\it Case I. \quad  In the absence of the differential gain or loss term $\mu(z)$, i.e., $\mu(z)\equiv 0.$}
\bes \label{ss} \bee
 \label{ss1} &&\tau(z,t)=\tau_1(z)t+\tau_0(z), \,\,\, \varphi(z,t)=\varphi_1(z)t+\varphi_0(z), \\
\label{ss2} && \eta(z)\!=\!\frac{\sqrt{2}}{4\nu}\!\int_0^z\!\!\alpha_1(s)\tau_1^3(s)ds,\,\,
       \rho(z)\!=\!\rho_0\!\exp\!\left[\!\int_0^z\!\gamma(s)ds\!\right]\!\!, \qquad\,\, \\
 \label{ss3} && \beta(z)=\alpha_1(z)\!\!\left[3\varphi_1(z)-\frac{\tau_1(z)}{2\sqrt{2}\nu}\!\right], \\
 \label{ss4} && g(z)=\frac{\alpha_1(z)\tau_1^2(z)}{\rho^2(z)}
      \left[3\varphi_1(z)+\frac{G\tau_1(z)}{2\sqrt{2}\nu}\right]\!\!,\\
\label{ss5} &&\sigma(z,t)\!=\!\frac{\dot{\tau_1}(z)t+\!\dot{\tau_0}(z)}{\tau_1(z)}
    +\alpha_1(z)\varphi_1(z)\!\!\left[\!\frac{\tau_1(z)}{\sqrt{2}\nu}-3\varphi_1(z)\!\right]\!\!, \,\,\,\\
 \label{ss6} && \nonumber V(z,t)=\left[\frac{\varphi_1(z)\dot\tau_1(z)}{\tau_1(z)}-\dot\varphi_1(z)\right]t
     +\frac{\varphi_1(z)\dot\tau_0(z)}{\tau_1(z)} \\
     && \qquad\quad +\alpha_1(z)\varphi_1^2(z)\left[\frac{\tau_1(z)}{2\sqrt{2}\nu}-\varphi_1(z)\right]-\dot\varphi_0(z), \\
\label{ss7}   &&  \alpha_3(z)=-\alpha_2(z)=-3\alpha_1(z)\tau_1^2(z)\rho^{-2}(z),
\ene\ees
where the dot over the variables denotes the space derivative,  $\tau_1(z)\not\equiv 0$ is the inverse of the widths of the pulse, $-\tau_0(z)/\tau_1(z)$ is the center of the pulse, $\varphi_j(z) (j=1,2)$ denote the frequency shift and the phase-front curvature, respectively, $\tau_j(z) (j=0,1)$, $\varphi_j(z) (j=1,2)$, $\gamma(z)$ and $\alpha_1(z)$ are free differentiable functions of space $z$, $\rho_0$ is a constant.

It is easy to see that for the absence of differential gain or loss term $\mu\equiv 0$, the GVD parameter $\beta(z,t)$, SPM parameter $g(z,t)$ and gain or loss term $\gamma(z,t)$ are only functions of space $z$. It follows from Eqs.~(\ref{ss2})-(\ref{ss4}) that the gain or loss
terms $\gamma(z)$ can be used to manipulate the
amplitude $\rho(z)$, SS parameter $\alpha_2(z)$, and SPM parameter $g(z)$. The TOD parameter $\alpha_1(z)$ is used to control the variable $\eta(z)$, the GVD parameter $\beta(z)$, SFS parameter $\alpha_3(z)$, SPM parameter $g(z)$, $\sigma(z)$, and potential $V(z,t)$.

Notice that the coefficients of
first degree term in $\tau(z,t)$ and the phase $\varphi(z,t)$ differs from the only case in which they must be constants~\cite{hrw} since we consider more two terms in Eq.~(\ref{chnls}), i.e., the group velocity term $\sigma(z,t)$ and the external potential $V(z,t)$. The varying parameters $\tau_1(z)$ and $\varphi_1(z)$ will excite complicated structures which may be useful to control the propagation of optical ultra-short pulses (see, e.g.,
\cite{p1,p2,p3,pg1,pg2} for the similar waves).

To clearly understand these constraints about the coefficients in Eq.~(\ref{chnls}) we will present the following  special case. If we require that all coefficients in Eq.~(\ref{chnls}) depend on only space, i.e., the external potential $V(z,t)$ and the group velocity $\sigma(z,t)$ depend on only space $z$, then it follows from
Eqs.~(\ref{ss5}) and (\ref{ss6}) that we have the conditions
 \bee
   \dot{\tau_1}(z)=0, \qquad   \dot{\varphi_1}(z)=0,
 \ene
i.e., $\tau_1(z)$ and $\varphi_1(z)$ should be real constants:
 \bee
  \tau_1(z)=\sqrt{2} C_1, \qquad  \varphi_1(z)=C_2, \ene
with $C_{1,2}$ being constants. In this case, system (\ref{ss}) becomes the simple form
\bes \label{ssg} \bee
 \label{ssg1} &&\tau(z,t)=C_1t+\tau_0(z), \,\,\, \varphi(z,t)=C_2t+\varphi_0(z), \\
\label{ssg2} && \eta(z)\!=\!\frac{C_1^3}{\nu}\!\int_0^z\!\!\!\!\alpha_1(s)ds,\,\,
 \beta(z)\!=\!\frac{(6\nu C_2\!\!-\!\!C_1)}{2\nu}\alpha_1(z), \\
 \label{ssg3} && g(z)=\frac{C_1^2(6\nu C_2-C_1G)}{\nu}\frac{\alpha_1(z)}{\rho^2(z)},\\
\label{ssg4} &&\sigma(z)=\frac{C_2(C_1-3\nu C_2)}{\nu}\alpha_1(z)+\frac{\dot{\tau_0}(z)}{\sqrt{2} C_1},\\
 \label{ssg5} && \nonumber V(z)=\frac{C_2^2(C_1-2\nu C_2)}{2\nu}\alpha_1(z)+\frac{C_2\dot\tau_0(z)}{\sqrt{2} C_1}-\dot\varphi_0(z), \\
\label{ssg6}   &&  \alpha_3(z)=-\alpha_2(z)=-6C_1^2\alpha_1(z)\rho^{-2}(z),
\ene\ees
and $\rho(z)$ is given by Eq.~(\ref{ss2}), where $\alpha_1(z)$,\, $\gamma(z)$,\, $\tau_0(z)$,\, $\varphi_0(z)$ are free functions of space, $C_1,\, C_2, \nu,\, \rho_0$ are free constants.

Notice that these free parameters can modulate these coefficients $\beta(z)$,\, $g(z)$,\, $\sigma(z)$,\, $V(z)$,\, $\alpha_{2,3}(z)$ in Eq.~(\ref{chnls}), which enlarge the scope of the coefficients of the study model (see, e.g.,~\cite{hnlse,hnlst,hnlsc, vy}) such that more wave phenomena may be generated by modulating these coefficients.

\vspace{0.1in}
{\it Case II. \quad  In the presence of the differential gain or loss term $\mu(z)$, i.e., $\mu(z)\not\equiv 0$.}
\bes \label{pa2} \bee
 \label{pa21} &&\tau=\tau_1(z)t+\tau_0(z), \,\,\,\, \eta=\frac{\sqrt{2}}{4\nu}\!\int_0^z\!\!\alpha_1(s)\tau_1^3(s)ds,\\
\label{pa22} && \varphi=\frac{\mu(z)}{6\alpha_1(z)}t^2+\varphi_1(z)t+\varphi_0(z), \\
\label{pa23} && \gamma=-\frac{\mu^2(z)}{3\alpha_1(z)}t+\gamma_0(z),\\
\label{pa24} && \rho\!=\!\rho_0\exp\left\{\int_0^z\left[\mu(s)\left(\varphi_{1}(s)\!-\!\frac{\tau_{1}(s)}{6\sqrt{2}\nu}\!\right)
  +\!\gamma_{0}(s)\!\right]\!\!ds\!\!\right\}\!,\qquad  \\
\label{pa25} && \beta=\mu(z)t+\alpha_1(z)\left[3\varphi_1(z)-\frac{\tau_1(z)}{2\sqrt{2}\nu}\!\right]\!\!,\\
\label{pa26} && g=\frac{\tau_1^2(z)}{\rho^2(z)}
      \left[\mu(z)t+3\alpha_1(z)\varphi_1(z)+\frac{G\alpha_1(z)\tau_1(z)}{2\sqrt{2}\nu}\right]\!\!,\\
\label{pa27} &&\sigma=-\frac{\mu^2(z)}{3\alpha_1(z)}t^2+\sigma_1(z)t+\sigma_0(z), \\
\label{pa28} &&  V=-\frac{\mu^3(z)}{27\alpha_1^2(z)}t^3+v_2(z)t^2+v_1(z)t+v_0(z),\\
\label{pa29} &&\alpha_3(z)=-\alpha_2(z)=-\frac{3\alpha_1(z)\tau_1^2(z)}{\rho^2(z)},
\ene\ees
 where $\varphi_j(z)\,(j=0,1)$, $\tau_j(z)\, (j=0,1)$, $\gamma_0(z)$ and $\alpha_1(z)$ are arbitrary differentiable functions of space $z$, $\rho_0$ is a constant. We have introduced these functions in Eqs.~(\ref{pa27}) and (\ref{pa28}) given by
 \bes\bee
 && \sigma_1(z)=\displaystyle\frac{\mu(z)}{6\nu}\!\!
     \left[\sqrt{2}\tau_1(z)-12\nu\varphi_1(z)\!\right]\!\!+\frac{\dot{\tau_1}(z)}{\tau_1(z)},\quad \vspace{0.08in} \\
 &&\sigma_0(z)=\displaystyle\frac{\alpha_1(z)\varphi_1(z)}{\sqrt{2}\nu}\!\!
      \left[\tau_1(z)-3\sqrt{2}\nu\varphi_1(z)\!\right]+\frac{\dot{\tau_0}(z)}{\tau_1(z)},\qquad
    \ene\ees
 \bee\begin{array}{l}
  v_2(z)=\displaystyle
      \frac{\mu^2(z)}{3\alpha_1(z)}\!\left[\!\frac{\tau_1(z)}{6\sqrt{2}\nu}-\varphi_1(z)\!\right]-\left[\frac{\mu(z)}{6\alpha_1(z)}\right]_z  \qquad\qquad \vspace{0.08in}\\
  \displaystyle\qquad\qquad\quad +\frac{\mu(z)\dot{\tau_1}(z)}{3\alpha_1(z)\tau_1(z)},
  \end{array} \ene
\bee\begin{array}{l}
    v_1(z)=\displaystyle
      \mu(z)\varphi_1(z)\!\left[\!\frac{\tau_1(z)}{3\sqrt{2}\nu}-\varphi_1(z)\!\right]-\dot{\varphi_1}(z)\qquad\qquad\quad \vspace{0.08in}\\
     \displaystyle\qquad\qquad +\frac{3\alpha_1(z)\dot{\tau_1}(z)\varphi_1(z)+\mu(z)\dot{\tau_0}(z)}{3\alpha_1(z)\tau_1(z)},\qquad
     \end{array} \ene
\bee\begin{array}{l}
 v_0(z)=\displaystyle
       \alpha_1(z)\varphi_1^2(z)\!\left[\!\frac{\tau_1(z)}{2\sqrt{2}\nu}-\varphi_1(z)\!\right]-\dot{\varphi_0}(z)
       \qquad \vspace{0.08in}\\
     \displaystyle\qquad\qquad\quad
       \!\!+\!\frac{\dot{\tau_0}(z)\varphi_1(z)}{\tau_1(z)},
          \end{array} \ene

It is easy to see that for the case of the presence of differential gain or loss term $\mu(z)\not\equiv 0$, the GVD parameter $\beta(z,t)$, SPM parameter $g(z,t)$ and gain or loss term $\gamma(z,t)$ are all functions of both $z$ and $t$ , which differ from the considered usual higher-order NLS equation with varying coefficients. The differential gain or loss term $\mu(z)$ can be used to modulate the phase, gain or loss term $\gamma(z,t)$, the amplitude $\rho(z)$, the GVD parameter $\beta(z,t)$, the SS parameter $\alpha_2(z)$, SPM parameter $g(z,t)$, $\sigma(z,t)$ and potential $V(z,t)$.
The phase $\varphi(z,t)$ is a second degree polynomial in $t$ with coefficients being functions of $z$ which is similar to one in the solutions of the NLS equation with varying coefficients, but the external potential $V(z,t)$ is a third degree polynomial in $t$ with coefficients being functions of $z$ which differs from one in the solutions of the NLS equation with varying coefficients~\cite{pg2}.

Notice that the solutions in Case I can be found directly from ones in Case II with $\mu\equiv 0$, but we here list Case I in order to clearly point out that GVD $\beta(z)$, SPM $g(z)$, and gain or loss $\gamma(z)$ depend only on space $z$, and $\sigma(z,t)$ and potential $V(z,t)$ are the linear functions of time with coefficients being functions of space in the absence of the differential gain or loss term $\mu$ where as GVD $\beta(z,t)$, SPM $g(z,t)$, and gain or loss $\gamma(z,t)$ are linear functions of time with coefficients being functions of space, and $\sigma(z,t)$ and potential $V(z,t)$ are the second-degree and third-degree functions of time with coefficients being functions of space in the present of the differential gain or loss term $\mu(z)$.

Thus these chosen differentiable functions in the amplitude $\rho(z)$, GVD $\beta(z,t)$, SPM $g(z,t)$, external potential $V(z,t)$, the gain or loss $\gamma(z,t)$, TOD $\alpha_1(z)$, SS $\alpha_2(z)$, SFS $\alpha_3(z)$, the phases $\varphi(z,t)$, and new variables $(\eta(z), \tau(z,t)$ can excite
abundant nonlinear wave structures of Eq.~(\ref{chnls}) such as period wave equations, multi-soliton solutions, and even rogue wave solutions.

Note that from the analytical conditions given by (\ref{ss}) and (\ref{pa2}), the
TOD parameter $\alpha_1(z)$ influences all system parameters, such
as GVD $\beta(z,t)$, SPM $g(z,t)$, external potential $V(z,t)$,
the gain or loss $\gamma(z,t)$,  SS
$\alpha_2(z)$, SFS $\alpha_3(z)$ and the form factors of
solutions, such as the amplitude $\rho(z)$ and the phase
$\varphi(z,t)$. Solutions found can exist only under certain
conditions and the system parameter functions can not be all chosen
independently. Thus, similar to the corresponding discussions in Refs.~\cite{nlsvc, a68,a69},
we can choose the equation parameters suitably to
investigate the dynamic behaviors for solutions of equation (\ref{chnls}).
The choice of periodic function for system parameters leads to
alternating regions of positive/negative values, which are
required for an eventual stability of solutions~\cite{nlsvc, a68,a69}. In the
next section, we will discuss the dynamical behaviors of
optical rogue wave solutions in some periodic dispersion
systems.

%%%%%%%%%%%%%%%%%%%%%%%%%%%%%%%%%%%%%%%%%%%%%%%%%%%%%
\begin{figure*}
\begin{center}
{\scalebox{0.55}[0.5]{\includegraphics{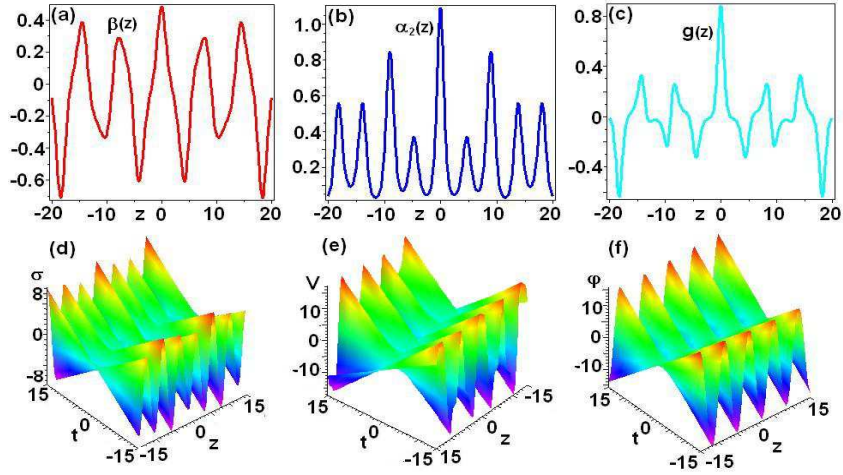}}}
\end{center}
\vspace{-0.25in} \caption{\small (color online). Profiles of (a) GVD $\beta(z)$  give by Eq.~(\ref{ss3});
(b) SS $\alpha_2(z)$ given by  Eq.~(\ref{ss7}); (c) SPM $g(z)$ given by Eqs.~(\ref{ss4});
(d) $\sigma(z,t)$ given by  Eq.~(\ref{ss5}); (e) the potential $V(z,t)$ given by  Eq.~(\ref{ss6}), and (f)
 the phase $\varphi(z,t)$ given by  Eq.~(\ref{ss1}) for the parameters given by  Eq.~(\ref{para1}) with
 $G=-1,\, k_1 = k_3=0.9,\, k_2 = 0.6,\, k_4 = 0.8,\, k_5=0.5,\, k_6=0.7,\, c_1=0.2$, and
$c_2=0.3$.} \label{fig:1}
\end{figure*}
%%%%%%%%%%%%%%%%%%%%%%%%%%%%%%%%%%%%%%%%%%%%%%%%%%%%

\section{Optical rogue wave solutions}

%%%%%%%%%%%%%%%%%%%%%%%%%%%%%%%%%%%%%%%%

Here we only consider optical rogue wave solutions of Eq.~(\ref{chnls}). To the aim, we need to know rogue wave solutions of
Eq.~(\ref{3nls}) with $G=-1$ which have been found~\cite{hnlsrw} in terms of the Darboux transformation
\bee
\Psi_{n+1}=\Psi_n-\frac{4iq_{n+1}p_{n+1}^{*}}{|p_{n+1}|^2+|q_{n+1}|^2} \,\,\, (n=0,1,2,...)\,\,
\ene
with the proper initial (`seed') solution $\Psi_0=e^{i\eta}$,  where $p_{n+1}^{*}$ denotes the complex conjugation of $p_{n+1}$, the characteristic functions $(p_n, q_n)$ and the solution $\Psi_n$ of Eq.~(\ref{3nls}) satisfy the simplified Lax pair in which the characteristic is chosen as $\lambda=i$)~\cite{hnlsrw}
\bes \label{dt}
 \bee
 && \left[\!\!\begin{array}{c} p_{n+1} \vspace{0.2in}\cr q_{n+1} \end{array}\!\!\right]_{\eta}
  =\left[\begin{array}{cc} -\frac{1}{\sqrt{2}} & \frac{i}{\sqrt{2}}\Psi^{*}_n \vspace{0.08in}\cr
      \frac{i}{\sqrt{2}}\Psi_n & \frac{1\lambda}{\sqrt{2}} \end{array} \right]
      \!\!\left[\!\!\begin{array}{c} p_{n+1} \vspace{0.2in}\cr q_{n+1}\end{array}\!\!\right], \\
 && \left[\!\!\begin{array}{c} p_{n+1} \vspace{0.2in}\cr q_{n+1}\end{array}\!\!\right]_{\tau}
 = \left[\!\!\begin{array}{cc} M_{11} & M_{12} \vspace{0.2in}\cr
     M_{21} & -M_{11} \end{array} \!\!\!\right]
      \!\!\!\left[\!\!\begin{array}{c} p_{n+1}\vspace{0.2in}\cr q_{n+1} \end{array}\!\!\right], \qquad
 \ene
 \ees
 where these components $M_{ij}$ are given by
 \bee
 \nonumber\begin{array}{l}
 M_{11}\!=\!\sqrt{2}\nu(\Psi^*\Psi_{\tau}-\Psi\Psi_{\tau}^{*})-(2+i/2)|\Psi|^2-4\nu-i, \vspace{0.08in}\cr
 \displaystyle M_{12}\!=\!2i\nu\Psi_{\tau\tau}^{*}\!-\!\frac{\Psi_{\tau}^{*}}{\sqrt{2}}(4i\nu-1)+\Psi^*(2i\nu|\Psi|^2\!+\!4i\nu\!-\!1),
 \vspace{0.08in} \cr
 \displaystyle M_{21}\!=\!2i\nu\Psi_{\tau\tau}\!+\!\frac{\Psi_{\tau}}{\sqrt{2}}(4i\nu-1)+\Psi(2i\nu|\Psi|^2\!+\!4i\nu\!-\!1),
\end{array}
 \ene
where $\Psi^{*}$ denotes the complex conjugation of the electric field $\Psi$.

In fact, on the basis of the above-obtained two invariant transformations of Eq.~(\ref{3nls}) given by Eqs.~(\ref{b1}) and {\ref{b2}), we can generate the generalized rogue wave solutions of Eq.~(\ref{3nls}) using its obtained rogue wave solutions~\cite{hnlsrw}. In the following as two representative examples, we
consider the low-order rational solutions of Eq.~(\ref{3nls})
which serve as prototypes of rogue waves to obtain the non-stationary rogue wave solutions of
Eq. (\ref{chnls}).

\subsection{First-order optical rogue wave solutions}

Here we consider the lowest order rational solutions of
Eq.~(\ref{3nls})~\cite{hnlsrw}, which serve as prototypes of
rogue waves. As a result, we obtain the first-order
self-similar (non-stationary) rogue wave solutions of Eq.~(\ref{chnls})
in the form
\bee \nonumber
 \psi_1(z,t)\!=\!\rho(z)\!\left\{\!\!1\!-\!\frac{4+8i\eta(z)}{1\!+\![\sqrt{2}\tau(z,t)\!+\!12\nu\eta(z)]^2\!+\!4\eta^2(z)}\right\} \\
\label{solu1} \times \exp[i\eta(z)+i\varphi(z,t)] \qquad\qquad\qquad
\ene
on the basis of the obtained similarity reduction transformation (\ref{trans}), where the new variables $\tau(z,t)$ and $\eta(z)$, the amplitude $\rho(z)$, and the phase $\varphi(z,t)$ are given by systems (\ref{ss})
and (\ref{pa2}). For the Case I, $\tau_j(z),\, \varphi_j(z)\, (j=0,1)$, $\alpha_1(z),\, \gamma(z)$ are used to modulate the non-stationary rogue wave solutions (\ref{solu1}) and coefficients of Eq.~(\ref{chnls}); whereas for Case II,  $\tau_j(z),\, \varphi_j(z)\, (j=0,1)$,\, $\alpha_1(z),\, \mu(z),\, \gamma_0(z)$ are used to control the non-stationary rogue wave solutions (\ref{solu1}) and coefficients of Eq.~(\ref{chnls}).

Here we simply analyze the physical mechanisms of rogue waves of Eq.~(\ref{chnls}) on the basis of the constraints in Section II. (i) for the case I about the constraints given by system ~(\ref{ss}), the TOD term $\alpha_1(z)$ in not equal to zero such that the SS $\alpha_2(z)$, and SFS parameter $\alpha_3(z)$ should not be zero. The GVD $\beta(z)$ and SPM $g(z)$ must not zero except for the case $\tau_1(z)=6\sqrt{2}\nu\varphi_1(z)$. Similarly, the group velocity $\sigma$ and external potential $V$ are zero or not. In brief, in this case given by  system ~(\ref{ss}), the TOD, SS and SFS terms are key to control Eq.~(\ref{chnls}) to generate rogue wave solutions. (ii) for the case II about the constraints given by system ~(\ref{pa2}), these coefficients in Eq.~(\ref{chnls}) are not all zero since $\mu(z)\not\equiv 0$. All these not-zero terms are modulated to make Eq.~(\ref{chnls}) admit the rogue wave solution (\ref{solu1}) with the phase $\varphi(z,t)$ givne by Eq.~(\ref{pa22}) being of the similar one as the NLS equation with varying coefficients~\cite{pg2}.

%%%%%%%%%%%%%%%%%%%%%%%%%%%%%%%%%%%%%%%%%%%%%%%%%%%%%
\begin{figure}[htp]
\begin{center}
{\scalebox{0.28}[0.25]{\includegraphics{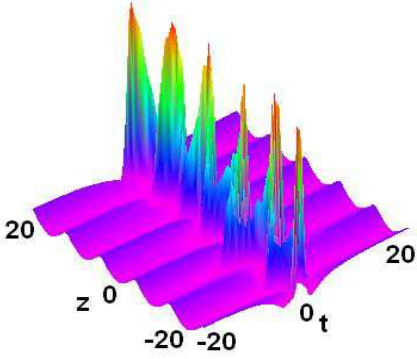}}}
\end{center}
\vspace{-0.2in} \caption{\small (color online). Profile of the wave intensity distribution
 $|\psi_1(z,t)|^2$ with ${\rm max}_{\{z, t\}}|\psi_1|^2\simeq 16.2$ defined by the solution (\ref{solu1})
for the parameters given by Eq.~(\ref{para1}) with
$k_1 = k_3=0.9,\, k_2 = 0.6,\, k_4 = 0.8,\, k_5=0.5,\, k_6=0.7,\, c_1=0.2$, and
$c_2=0.3$.} \label{fig:2}
\end{figure}
%%%%%%%%%%%%%%%%%%%%%%%%%%%%%%%%%%%%%%%%%%%%%%%%%%%%
%%%%%%%%%%%%%%%%%%%%%%%%%%%%%%%%%%%%%%%%%%%%%%%%%%%%%
\begin{figure}[htp]
\begin{center}
{\scalebox{0.28}[0.3]{\includegraphics{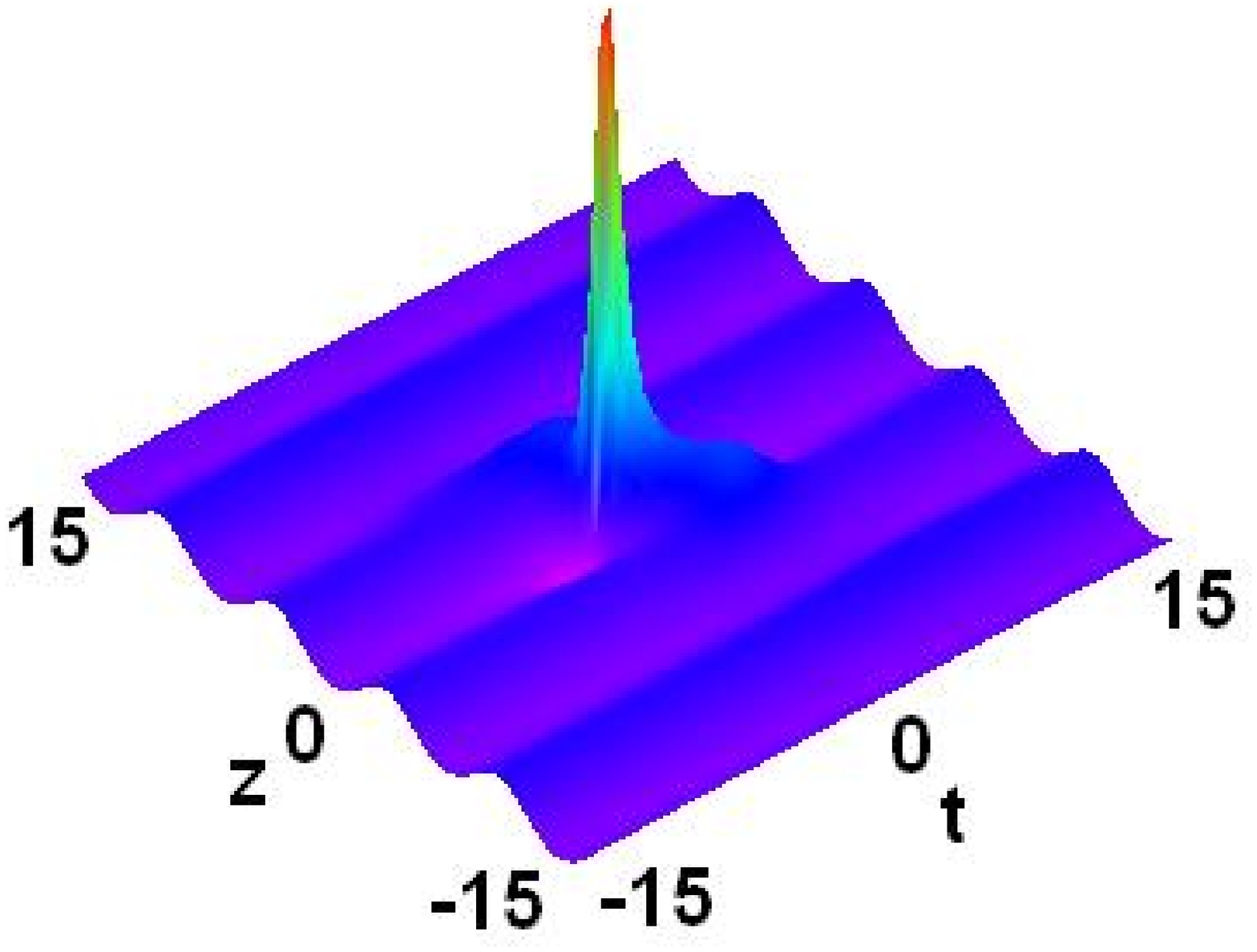}}}
\end{center}
\vspace{-0.15in} \caption{\small (color online). Profile of the wave intensity distribution
 $|\psi_1(z,t)|^2$ with ${\rm max}_{\{z, t\}}|\psi_1|^2\simeq 6.4$ defined by the solution (\ref{solu1})
for the parameters given by Eq.~(\ref{para2}) with
$k_4 =0.5$ and $c_2=0.1$.} \label{fig:3}
\end{figure}
%%%%%%%%%%%%%%%%%%%%%%%%%%%%%%%%%%%%%%%%%%%%%%%%%%%%

%%%%%%%%%%%%%%%%%%%%%%%%%%%%%%%%%%%%%%%%%%%%%%%%%%%%%
\begin{figure*}[htp]
\begin{center}
{\scalebox{0.65}[0.65]{\includegraphics{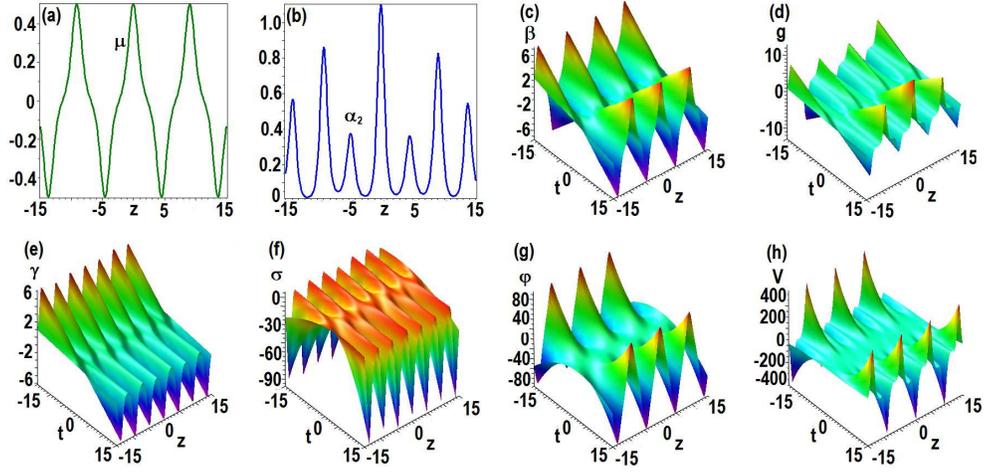}}}
\end{center}
\vspace{-0.15in} \caption{\small (color online). Profiles of (a) the differential gain or loss term $\mu(z)$ given by Eq.~(\ref{pcase2});
 (b) SS $\alpha_2(z)$ given by Eq.~(\ref{pa29}); (c) GVD $\beta(z,t)$  give by Eq.~(\ref{pa25}); (d)  SFS $g(z,t)$ given by Eqs.~(\ref{pa26}); (e) the gain or loss term $\gamma(z,t)$ given by Eq.~(\ref{pa23}); (f) $\sigma(z,t)$ given by  Eq.~(\ref{pa27}); (g) the phase $\varphi(z,t)$ given by  Eq.~(\ref{pa22}); (h)  the potential $V(z,t)$ given by  Eq.~(\ref{pa28}) for the parameters given by  Eq.~(\ref{pcase2}) with
 $G=-1,\, k_1 = k_3=0.9,\, k_2 = 0.6,\, k_4 = 0.8,\, k_5=0.9,\, k_6=0.7,\, c_1=0.2,\, c_2=0.3$, and
$\mu_0=0.5$.} \label{fig:4}
\end{figure*}
%%%%%%%%%%%%%%%%%%%%%%%%%%%%%%%%%%%%%%%%%%%%%%%%%%%%

To make sure that these TOD $\alpha_1(z)$, \, SS $\alpha_2(z)$, SFS $\alpha_3(z)$,\, GVD $\beta(z)$ for Case I (or the coefficient of GVD $\beta(z,t)$ in time for Case II), SPM $g(z)$ for Case I (or the coefficient of SPM $g(z,t)$ in time for Case II), $\gamma(z)$ for Case I (or the coefficient of $\gamma(z,t)$ in time for Case II), the coefficients of $\sigma(z,t)$ in time, and the coefficients of the external potential $V(z,t)$ in time in system (\ref{chnls}) are bounded for realistic cases.

For the illustrative purposes, we choose these free parameters in Case I in the form
\bee \label{para1}
\begin{array}{l}
 \rho_{0}=1.0, \,\,  \nu=0.6, \vspace{0.05in}\cr
 \tau_1(z)={\rm dn}(z, k_1),\,\, \tau_0(z)={\rm cn}(z,k_2), \vspace{0.05in}\cr
 \alpha_1(z)=c_1{\rm dn}(z,k_3), \,\, \gamma(z)=c_2{\rm sn}(z,k_4){\rm dn}(z,k_4),\vspace{0.05in}\cr
 \varphi_1(z)={\rm sn}(z, k_5),\,\, \varphi_0(z)={\rm cn}(z, k_6)
\end{array}
\ene
where sn, cn, and dn stand for the respective
Jacobi elliptic functions, and $k_j\in (0,1)\, (j=1,2,...,6)$ are their moduli, and $c_j\,(j=1,2)$ are real-valued constants.

Figure~\ref{fig:1} depicts the profiles of GVD $\beta(z)$,  SS $\alpha_2(z)$, SPM $g(z)$ given by system (\ref{ss}) vs space $z$ and the profiles of $\sigma(z,t),\, V(z,t),\, \varphi(z,t)$ given by system(\ref{ss}) for the parameters are given by Eq.~(\ref{para1}). The evolution
of intensity distributions $|\psi_1(z,t)|^2$ of the rogue wave
fields given by Eq.~(\ref{solu1}) is illustrated in Fig.~\ref{fig:2}.
We can see that the solutions are localized in
time and keep the localization infinitely in space, which are generated arising from the varying coefficients and
differ from the usual rogue wave solutions (see, e.g.,~\cite{PS, na, hnlsrw,frw,frw2}). The solutions may be useful for the experimentalists to modulate these coefficients to generate different rogue wave phenomenon in nonlinear optics.

On the other hand, if we choose the free parameters in
the another form
\bee\label{para2}
\begin{array}{l}
\tau_1(z)=1+0.1\sin(z),\,\,\, \tau_0(z)=\cos(z), \vspace{0.05in}\cr
 \alpha_1(z)=0.2+0.1\sin(z),\,\, \nu=0.1,
\end{array}
\ene
and $\rho_0,\, \gamma(z)$, and $\varphi_j(z)\, (j=0,1)$ are same as the ones
given by Eq.~(\ref{para1}), then the evolution of intensity distribution
of the rogue wave solutions (\ref{solu1}) will
be changed. Figure~\ref{fig:3} displays the profile
of the rogue wave solution (\ref{solu1}). The solutions are localized
both in time and almost in space thus almost revealing the usual
rogue wave features (see, e.g., \cite{hnlsrw,na}).

For Case II, we choose the different gain or loss term
\bee \label{pcase2}
\begin{array}{l}
\mu(z)=\mu_0{\rm cn}(z,k_7){\rm dn}(z,k_7), \vspace{0.05in}\cr
\gamma_0(z)=c_3{\rm sn}(z,k_8){\rm dn}(z,k_8),
\end{array}\ene
where $c_j\,(j=3,4)$ are real-valued constants.

%%%%%%%%%%%%%%%%%%%%%%%%%%%%%%%%%%%%%%%%%%%%%%%%%%%%%
\begin{figure}
\begin{center}
{\scalebox{0.3}[0.3]{\includegraphics{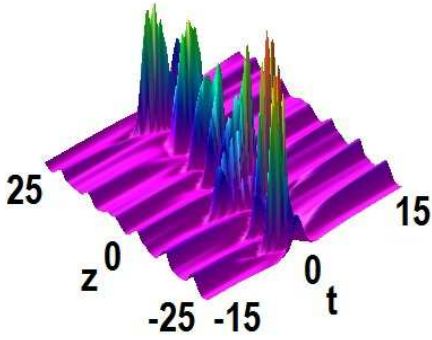}}}
\end{center}
\vspace{-0.15in} \caption{\small (color online). Profile of the wave intensity distribution
 $|\psi_1(z,t)|^2$ with ${\rm max}_{\{z, t\}}|\psi_1|^2\simeq 30.7$ defined by the solution (\ref{solu1})
for the parameters given by Eqs.~(\ref{pcase2}) and (\ref{para1}) with $k_1 = k_3=0.9,\, k_2 = 0.6,\, k_4 = 0.8,\, k_5=0.9,\, k_6=0.7,\, c_1=0.2,\, c_2=0.3$, and
$\mu_0=0.5$. } \label{fig:5}
\end{figure}
%%%%%%%%%%%%%%%%%%%%%%%%%%%%%%%%%%%%%%%%%%%%%%%%%%%%

Figure~\ref{fig:4} depicts the profiles of $\mu(z)$,  GVD $\beta(z,t)$,  SS $\alpha_2(z)$, SPM $g(z,t),$  $\sigma(z,t),\, V(z,t),\, \varphi(z,t)$ given by system(\ref{pa2}) for the parameters are given by Eqs.~(\ref{pcase2}) and (\ref{para1}).
The evolution
of intensity distributions ($|\psi_1(z,t)|^2$) of the rogue wave
fields given by Eq.~(\ref{solu1}) with parameters given by Eqs.~(\ref{pcase2}) and (\ref{para1}) is illustrated in Fig.~\ref{fig:5}.

%%%%%%%%%%%%%%%%%%%%%%%%%%%%%%%%%%%%%%%%%%%%%%%%%%%%%
\begin{figure}
\begin{center}
{\scalebox{0.28}[0.26]{\includegraphics{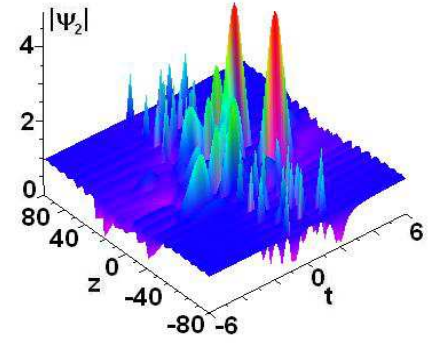}}}
\end{center}
\vspace{-0.1in} \caption{\small (color online). Profile of the amplitude distribution
 $|\psi_2(z,t)|$ with ${\rm max}_{\{z, t\}}|\psi_2|\simeq 4.9$ defined by the solution (\ref{solu2})
for the parameters given by Eq.~(\ref{para1}) with
$k_1=k_3=0.9,\, k_2 = 0.6,\, k_4 = 0.8,\, k_5=0.5,\, k_6=0.7,\, c_1=0.2$, and
$c_2=0.03$ except for $\nu=6.0$.} \label{fig:6}
\end{figure}
%%%%%%%%%%%%%%%%%%%%%%%%%%%%%%%%%%%%%%%%%%%%%%%%%%%%

\subsection{Second-order optical rogue wave solutions}

Here we consider the interaction of two rogue waves of Eq.~(\ref{chnls}). When a second-order rogue wave solution of the higher-order Hirota equation (\ref{3nls})~\cite{hnlsrw} is used to the obtained similarity transformation (\ref{trans}), we can obtain the second-order self-similar rogue wave solutions of Eq.~(\ref{chnls}) in the form
\bee\label{solu2}
\psi_2(z,t)\!=\!\rho(z)\!\left[1\!-\!\frac{P(\eta,\tau)\!+\!i\eta Q(\eta,\tau)}{H(\eta,\tau)}\right]
  \!e^{i[\eta+\varphi(z,t)]}, \qquad
\ene
where these functions $P(\eta,\tau),\, Q(\eta,\tau)$ and $H(\eta,\tau)$ are given by
\bee
\nonumber\begin{array}{l}
 P\!=\! 48\tau^4+1152\sqrt{2}\nu \eta\tau^3+144\tau^2[4\eta^2(36\nu^2+1)+1] \quad \vspace{0.08in}\cr
\quad\,\,\,\, +576\sqrt{2}\nu\eta\tau[12\eta^2(12\nu^2+1)+7]-36  \vspace{0.08in}\cr
\quad\,\,\,\, +192\eta^4[216\nu^2(6\nu^2+1)+5]\!+\! 864\eta^2(44\nu^2+1),\quad
  \end{array}
\ene
\bee
\nonumber\begin{array}{l}
 Q\!=\!96\tau^4+2304\sqrt{2}\nu \eta\tau^3+96\tau^2[4\eta^2(108\nu^2+1)-3]\quad\quad \vspace{0.08in}\cr
\quad\,\,\,\, +1152\sqrt{2}\nu\eta\tau[4\eta^2(36\nu^2+1)+1]-360 \vspace{0.08in}\cr
\quad\,\,\,\,  +384\eta^4(36\nu^2+1)^2+192\eta^2(108\nu^2+1),\quad \end{array}
\ene
\bee
\nonumber\begin{array}{l}
 H\!\!=\!8\tau^6+288\sqrt{2}\nu\eta\tau^5+12\tau^4[4\eta^2(180\nu^2+1)+1] \quad\qquad \vspace{0.08in}\cr
 \quad\,\,\,  +96\sqrt{2}\nu\eta\tau^3[12\eta^2(60\nu^2+1)-1] \vspace{0.08in}\cr
 \quad\,\,\,    +6\tau^2\!\big\{\!16\eta^4\![216\nu^2(30\nu^2\!+\!1)\!+\!1]\!-\!24\eta^2(60\nu^2\!+\!1)\!+\!9\!\big\}\quad \vspace{0.08in}\cr
\quad\,\,\, +72\sqrt{2}\nu\eta\tau[16\eta^4(36\nu^2\!+\!1)^2\!-\!8\eta^2(108\nu^2\!-\!1)\!+\!17] \vspace{0.08in}\cr
\quad\,\,\, +64\eta^6(36\nu^2+1)^3-432\eta^4(52\nu^2+1)(12\nu^2-1) \vspace{0.08in}\cr
\quad\,\,\, +36\eta^2(556\nu^2+11)+9,
\end{array}
\ene
in which the new variables $\tau(z,t)$ and $\eta(z)$, the amplitude $\rho(z)$, and the phase $\varphi(z,t)$ are given by systems (\ref{ss}) or (\ref{pa2}), and $\nu$ is a real-valued constant. These parameters $\alpha_1,\, \alpha_2,\, \mu,\, \tau_1,\, \tau_0,\, \gamma,\,\phi_1,\,\phi_0,$ and $\nu$ can be used to control the wave propagation of second-order self-similar rogue wave solutions (\ref{solu2}) and the coefficients of Eq.~(\ref{chnls}).

%%%%%%%%%%%%%%%%%%%%%%%%%%%%%%%%%%%%%%%%%%%%%%%%%%%%%
\begin{figure}[!htp]
\begin{center}
{\scalebox{0.28}[0.26]{\includegraphics{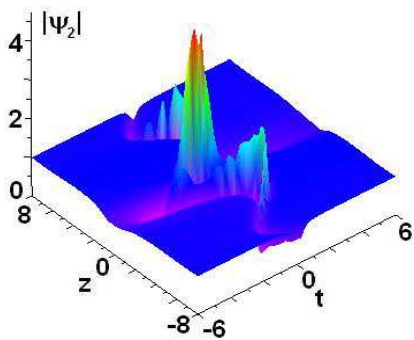}}}
\end{center}
\vspace{-0.1in} \caption{\small (color online). Profile of the amplitude distribution
 $|\psi_2(z,t)|$ with ${\rm max}_{\{z, t\}}|\psi_2|\simeq 4.7$ defined by the solution (\ref{solu2})
for the parameters given by Eq.~(\ref{para2}) with $k_4 = 0.5$ and
$c_2=0.02$ except for $\nu=5.0$} \label{fig:7}
\end{figure}
%%%%%%%%%%%%%%%%%%%%%%%%%%%%%%%%%%%%%%%%%%%%%%%%%%%%

%%%%%%%%%%%%%%%%%%%%%%%%%%%%%%%%%%%%%%%%%%%%%%%%%%%%
\begin{figure}[!htp]
\vspace{0.2in}
\begin{center}
{\scalebox{0.28}[0.26]{\includegraphics{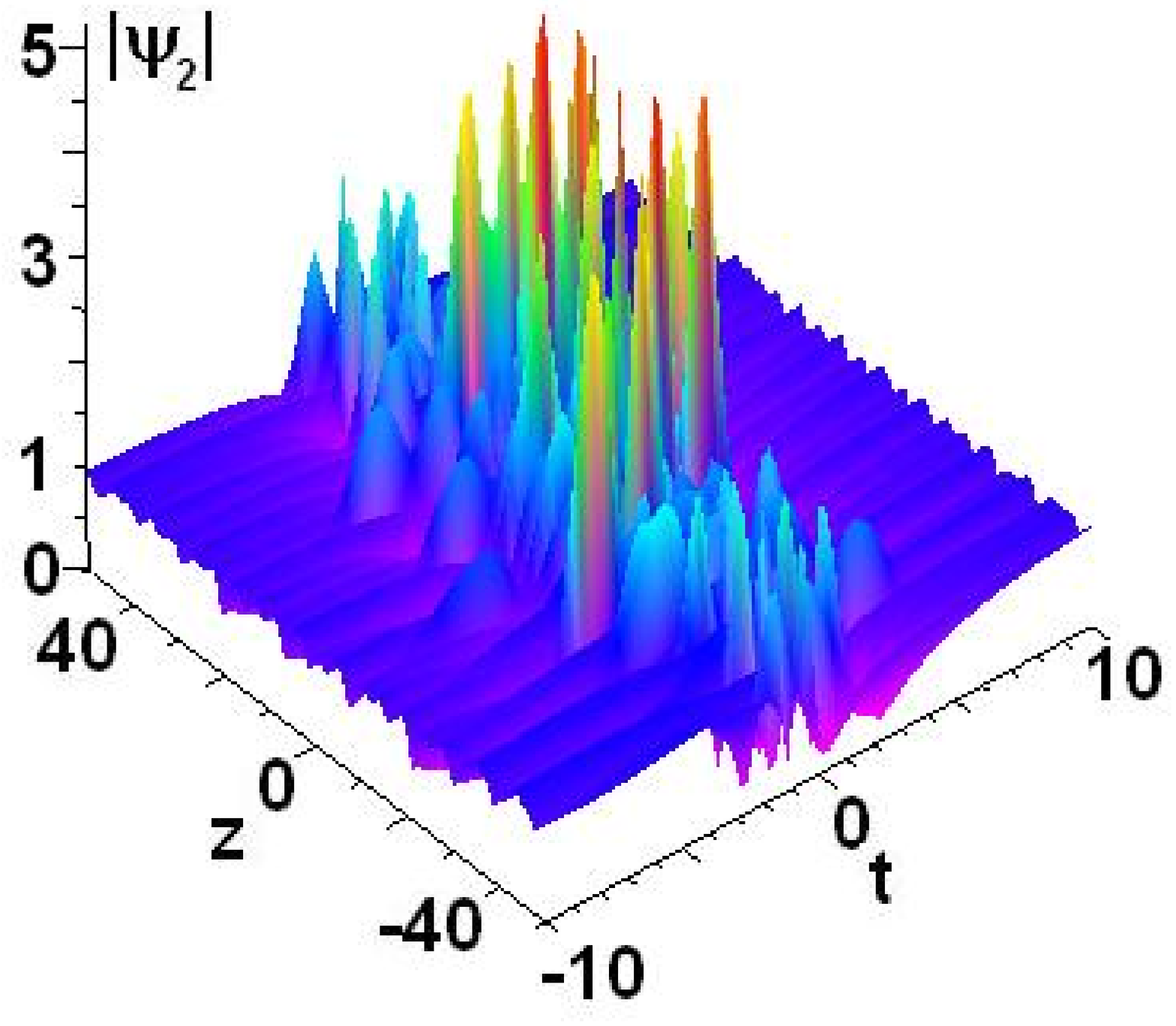}}}
\end{center}
\vspace{-0.1in} \caption{\small (color online). The profile of the amplitude distribution
 $|\psi_2(z,t)|$ with ${\rm max}_{\{z, t\}}|\psi_2|\simeq 5.3$ defined by the solution (\ref{solu2})
for the parameters given by Eq.~(\ref{pcase2}) with $k_1 = k_3=0.9,\, k_2 = 0.6,\, k_4 = 0.8,\, k_5=0.9,\, k_6=0.7,\, c_1=0.2,\, c_2=0.05$, and $\mu_0=0.01$ except for $\nu=2.0$.} \label{fig:8}
\end{figure}
%%%%%%%%%%%%%%%%%%%%%%%%%%%%%%%%%%%%%%%%%%%%%%%%%%%%

%%%%%%%%%%%%%%%%%%%%%%%%%%%%%%%%%%%%%%%%%%%%%%%%%%%%%
\begin{figure}[htp]
\begin{center}
{\scalebox{0.45}[0.45]{\includegraphics{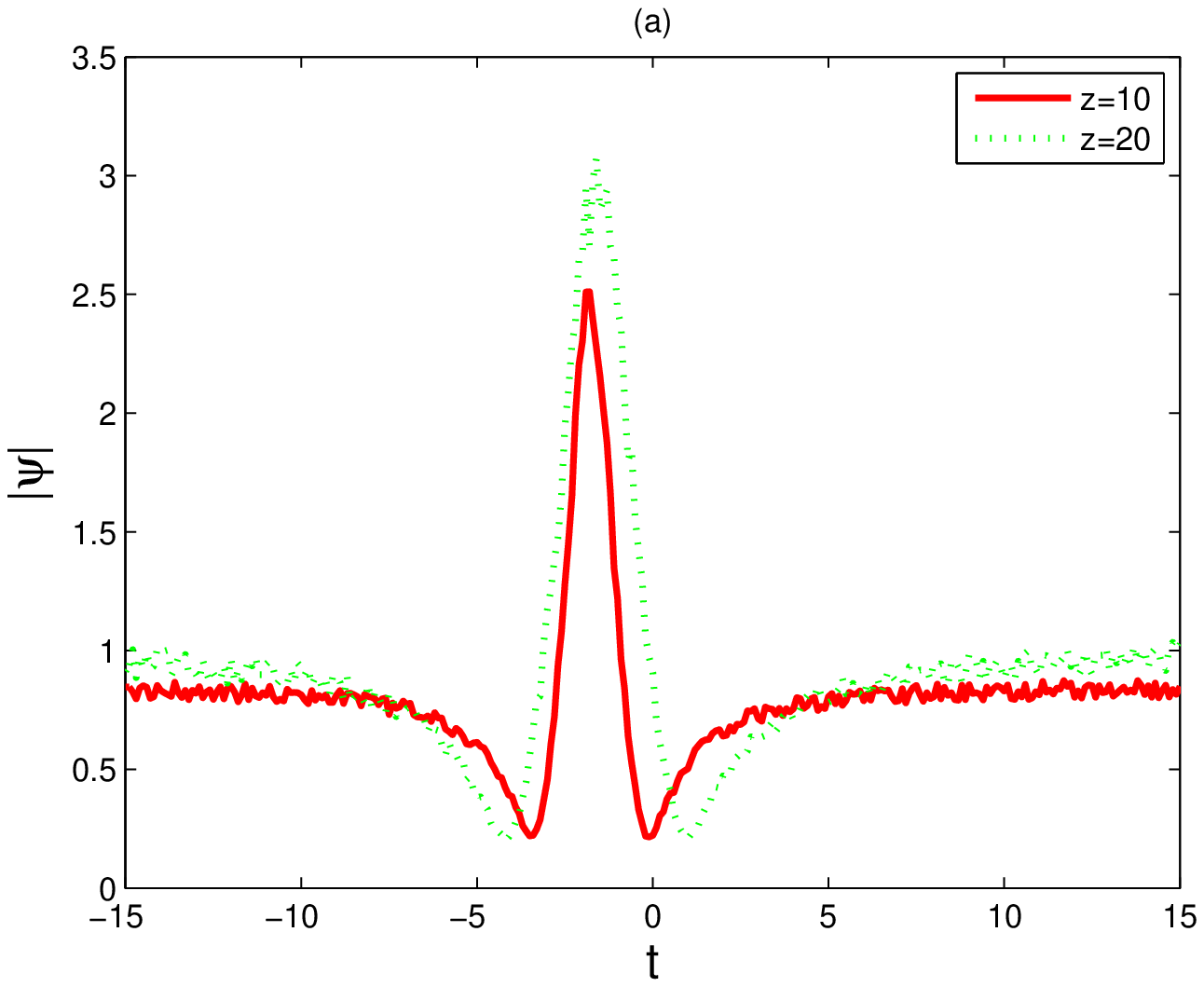}}}\\
{\scalebox{0.45}[0.45]{\includegraphics{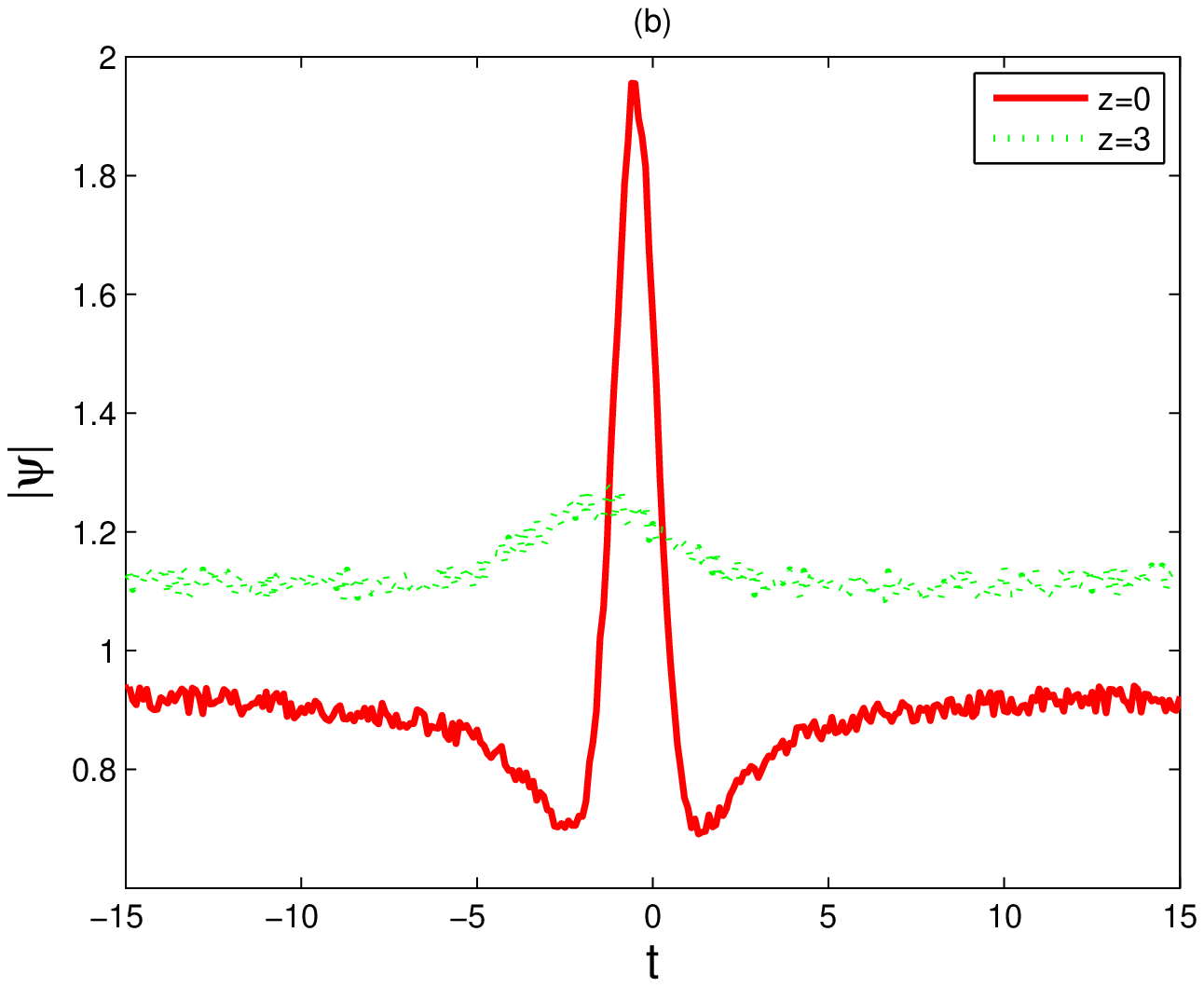}}}
\end{center}
\vspace{-0.1in} \caption{\small (color online). (a) and (b)
Numerical simulations of solution (\ref{solu1}) in the system (\ref{para1}) and
(\ref{para2}), respectively. The parameters are respectively the same as
that in Figs.~\ref{fig:2} and \ref{fig:3}, except for $k_2=k_4=k_5=k_6=0.9$.}\label{fig:9}
\end{figure}
%%%%%%%%%%%%%%%%%%%%%%%%%%%%%%%%%%%%%%%%%%%%%%%%%%%%

Similar to the subsection III.A, we choose three kinds of parameters given by Eqs.~(\ref{para1})-(\ref{pcase2}) to study the wave propagation of the second-order self-similar rogue wave solutions (\ref{solu2}).
Figures~\ref{fig:6}-\ref{fig:8} illustrate the amplitude distributions of the interactions of the second-order rogue wave solution (\ref{solu2}) in $(z,t)$-space for the chosen parameters. It is easy to find that the amplitude propagation of the physical field in this case is more complicated. Figure~\ref{fig:6} shows that the rogue wave solution is localized in time and not localized in space$z$ since the modulated parameters are some periodic functions of space $z$. In the another case, Figure~\ref{fig:7} illustrates that the  rogue wave solution is localized both in space $z$ and in time, however the complicated structure differs from the usual rogue wave feature, in particular one near the origin. For the presence of the differential gain or loss term $\mu(z)\not=0$, Figure~\ref{fig:8} illustrates that the self-similar rogue wave solution is localized both in space $z$ and in time, This may be useful to generate novel experimental results. Moreover, there may exist other integrable conditions making Eq.~(\ref{chnls}) to generate its other types of self-similar optical rogue wave solutions.

At last, we analyze the stability of some  obtained analytical solutions, that
is, how they evolve along distance when they are disturbed from
their analytically given forms. We perform direct numerical
simulations for Eq.~(\ref{chnls}) with initial fields coming from solution
(\ref{solu1}) in some cases. Two examples of such behaviors are displayed
in Fig.~\ref{fig:9}, which essentially presents a numerical rerun of Figs.~\ref{fig:2}
and \ref{fig:3}. Our preliminary results indicate no collapse, instead the
stable propagation are observed, except for some oscillations in
the wing of waves. Fig.~\ref{fig:9}(a) shows the stable propagation of
solution (\ref{solu1}) in the system (\ref{para1}). Fig.~\ref{fig:9}(b) exhibits the stable
wave shape of solution (\ref{solu1}) in the system (\ref{para2}). The comparison of
Figs.~\ref{fig:9}(a) and \ref{fig:9}(b) shows that solution (\ref{solu1}) in the system (\ref{para1}) is
more stable than solution (\ref{solu1}) in the system (\ref{para2}).

\section{conclusions}

In conclusion, we have systematically presented a proper transformation reducing the generalized higher-order nonlinear Schr\"odinger equation with varying coefficients to the Hirota equation with constant coefficients under some modulated coefficients. This self-similarity transformation and some constraints about the coefficients allows us to find certain class of exact self-similar solutions of the generalized HONLS equation (\ref{chnls}) in terms of the integrable Hirota equation (\ref{3nls}). In this paper we illustrated
 the approach on the basis of two lowest-order rogue wave solutions of the Hirota equation as seeding solutions
to study rogue wave solutions of Eq.~(\ref{chnls}), which are of complicated structures localized in time. For some chosen functions,
we study the wave propagations of rogue wave solutions. Moreover, we study the stability of the obtained first-order rogue wave solutions.

The used technique can be extended to investigate self-similar rogue waves of the coupled higher-order NLS equations with varying coefficients such as the external potentials, GVD, SPM, SS, SFS, and the gain or loss terms. The obtained self-similar rogue wave solutions may raise the possibility of relative experiments and potential applications in self-similar manner in nonlinear optics and other related fields of nonlinear science.

\acknowledgments
The work was supported by the NSFC under Grant Nos. 11071242, 61178091, and 11005092, and
the Zhejiang Provincial Natural Science Foundation of China under Grant No. Y13F050037.

%%%%%%%%%%%%%%%%%%%%%%%%%%%%%%%%%%%%%%%%%%%%%%%%%%%%%%%%%%%%%%

%\begin{thebibliography}{99}

%\end{thebibliography}

%%%%%%%%%%%%%%%%%%%%%%%%%%%%%%%%%%%%%%%%%%%%%%%%%%%%%%%%%%%%%%


\begin{references}

\bibitem{op} Kivshar Y S and Agrawal G P, {\em Optical Solitons: from
Fibers to Photonic Crystals} (Academic Press, New York,
2003).

\bibitem{op2}  Malomed B A, Mihalache D, Wise F, and
Torner L 2005 Spatiotemporal optical solitons {\it J. Opt. B: Quantum Semiclassical Opt.} {\bf 7}, 53R.

\bibitem{op3}  Hasegawa A and Kodama Y, Solitons
in Optical Communications (Oxford University Press, Oxford,
1995),

\bibitem{op4} Nie W J 1993 Optical nonlinearity: phenomena, applications, and materials {\it Adv. Mat.} {\bf 7/8} 520.


\bibitem{bec} Pitaevskii L and Stringari S {\em  Bose-Einstein Condensation}
(Oxford University Press, Oxford, 2003).

\bibitem{bec2} Pethick C J and Smith H {\em Bose-Einstein Condensation in
Dilute Gases} (Cambridge University Press, Cambridge, 2002).

\bibitem{bec3} Kartashov Y V, Malomed B A and Torner L 2011 Solitons in nonlinear lattices {\it Rev. Mod. Phys.} {\bf 83} 247.

\bibitem{bec4} Carretero-Gonz\'alez R, Frantzeskakis D J and Kevrekidis P G 2008 Nonlinear waves in Bose-Einstein condensates: physical relevance and mathematical techniques {\it Nonlinearity} {\bf 21} R139.

\bibitem{soliton} Ablowitz M J and Clarkson P A {\em Solitons, Nonlinear Evolution Equations and Inverse Scattering} (Cambridge, Cambridge University Press, 1991).

\bibitem{ch} Chen H H and Liu C S 1976 Solitons in nonuniform media {\it Phys. Rev. Lett.} {\bf 37} 693.

\bibitem{sl} Dudley J M {\it et al} 2007 Self-similarity in ultrafast nonlinear
optics {\it Nature Phys.} {\bf 3} 597-603.

\bibitem{sl2} Segev M, Soljacic M and Dudley J M 2012 Fractal optics and
beyond {\it Nature Photon.} {\bf 6} 209-10.

\bibitem{sl3} Hammani K, Boscolo S and  Finot C 2013 Pulse transition to similaritons in
normally dispersive fibre amplifiers {\it J. Opt.} {\bf 15} 025202-7.


\bibitem{nlsvc} Serkin V N and Hasegawa A 2000 Novel soliton solutions of the nonlinear Schr\"odinger equation model {\it Phys. Rev. Lett.}
 {\bf 85} 4502.

\bibitem{nlsvc1} Serkin V N and Hasegawa A 2000 Soliton management in the nonlinear Schr\"odinger equation model with varying dispersion, nonlinearity, and gain {\it JETP Letters} {\bf 72} 89¨C92.

\bibitem{nlsvc2} Kruglov V I, Peacock A C and Harvey J D 2003 Exact
self-similar solutions of the generalized nonlinear Schr\"odinger equation with distributed coefficients {\it Phys.
Rev. Lett.} {\bf 90} 113902.

\bibitem{pg1} Ponomarenko S A and Agrawal G P 2006 Do solitonlike self-similar waves exist in nonlinear optical media? {\it Phys. Rev. Lett.}  {\bf 97} 013901.

\bibitem{self1} Barenblatt G I Scaling, Self-Similarity, and Intermediate Asymptotics (Cambridge University Press, Cambridge 1996).

\bibitem{self2} Drazin P G and Jonson R S, Solitons: an Introduction (Cambridge University Press, 1988).


\bibitem{self3} Calogero F and Degasperis A, Spectral trasform and solitons (North Holland, Amsterdam, 1982).

\bibitem{self4} Lamb G L, Elements of soliton theory (John Wiley \& Sons Inc  1980).


\bibitem{p1} Belmonte-Beitia J {\it et al} 2008 Localized nonlinear waves in systems with time- and space-modulated nonlinearities {\it Phys. Rev. Lett.} {\bf 100} 164102.


\bibitem{p2} Kundu A 2009 Integrable nonautonomous nonlinear Schr\"odinger equations are equivalent
to the standard autonomous equation {\it Phys. Rev. E} {\bf 79} 015601R.

\bibitem{p3} Serkin V N, Hasegawa A and Belyaeva T L 2007 Nonautonomous Solitons in External Potentials, {\it Phys. Rev. Lett.}
 {\bf 98} 074102.

\bibitem{pg2} Yan Z Y 2010 Nonautonomous rogons in the inhomogeneous nonlinear Schr\"odinger equation with variable coefficients {\it Phys. Lett. A} {\bf 374} 672-679.

\bibitem{3d1} P\'erez-Garc\'ia V M,  Torresb P J and Konotop V V 2006 Similarity transformations for nonlinear Schr\"odinger equations with time-dependent coefficients {\it Physica D} {\bf 221} 31-36.

\bibitem{3d2} Beli\'c M {\it et al} 2008 Analytical Light Bullet Solutions to the Generalized  (3+1)-Dimensional
Nonlinear Schr\"odinger Equation {\it Phys. Rev. Lett.} {\bf 101}, 123904.

\bibitem{3d3} Yan Z Y and Konotop V V 2009 Exact solutions to three-dimensional generalized nonlinear Schr\"odinger equations with varying
potential and nonlinearities {\it Phys. Rev. E} {\bf 80} 036607.


\bibitem{nls3a} Kodama Y 1985 Optical Solitons in a Monomode Fiber {\it J. Stat. Phys.} {\bf 39} 597-18.

\bibitem{nls3b} Kodama Y and Hasegawa A 1987 Nonlinear pulse propagation in a monomode dielectric guide {\it IEEE J. Quantum Electron.}
  {\bf 23} 510-15.

\bibitem{nlsh} Hirota R 1973 Exact envelope-soliton solutions of a nonlinear wave equation {\it J. Math. Phys.} {\bf 14} 805-5.
\bibitem{nlss} Sasa N and Satsuma J 1991 New-type soliton solution for a higher-order nonlinear Schr\"odinger equation {\it  J. Phys. Soc. Jpn.} {\bf 60}  409-9.

\bibitem{nlsk} Kaup D J and Newell A C 1978 An exact solution for a derivative nonlinear Schr\"odinger equation {\it J. Math. Phys.} {\bf 19} 798.

\bibitem{nlsc} Chen H H, Lee L C and Lim C S 1979 Integrability of nonlinear hamiltonian systems
by inverse scattering method {\it Phys. Scr.} {\bf 20} 490-492.


\bibitem{hnlsc} Gedalin M, Scott T C and Band Y B 1997 Optical Solitons in the Higher Order Nonlinear Schr\"odinger Equation {\it Phys. Rev. Lett.} {\bf 78} 448.

\bibitem{li} Li Z H {\it et al} 2000 New Types of Solitary Wave Solutions for the Higher Order Nonlinear Schr\"odinger Equation {\it Phys. Rev. Lett.}
{\bf 84} 4096.

\bibitem{gi} Gilson C 2003 Sasa-Satsuma higher-order nonlinear Schro¡§dinger equation and its bilinearization
and multisoliton solutions {\it Phys. Rev. E} {\bf 68} 016614.

 \bibitem{vy} Vyas V M {\it et al} 2008 Chirped chiral solitons in the nonlinear Schr\"odinger equation with self-steepening and self-frequency shift {\it Phys. Rev. E} {\bf 78} 021803R.

\bibitem{hc1} Hao R Y, Li L, Li Z H and Zhou G S 2004 Exact multisoliton solutions of the higher-order nonlinear Schr\"odinger equation with variable coefficients {\it Phys. Rev. E} {\bf 70} 066603.

\bibitem{hc2} Zhang J F, Yang Q and Dai C Q 2005 Optical quasi-soliton solutions for higher-order nonlinear Schr\"odinger equation with variable coefficients {\it Opt. Commun.} {\bf 248} 257-265.

\bibitem{hc3} Wang J F {\it et al} 2006 Generation, compression and propagation of pulse trains under higher-order effects  {\it Opt. Commun.} {\bf 263} 328-336.

\bibitem{hc4} Li J {\it et al} 2007 Soliton-like solutions of a generalized variable-coefficient higher order nonlinear Schr\"odinger equation from inhomogeneous optical fibers with symbolic computation {\it J. Phys. A: Math. Theor.} {\bf 40} 13299-13309.

\bibitem{hc5} Porsezian K {\it et al} 2007 Dispersion and nonlinear management for femtosecond optical solitons {\it Phys. Lett. A} {\bf 361} 504-508.

\bibitem{hnlse} Colman P {\it et al} 2010 Temporal solitons and pulse compression in
photonic crystal waveguides {\it Nature Photonics} {\bf 4} 862.

\bibitem{hnlst} Bhat N A R and Sipe J E 2001 Optical pulse propagation in nonlinear photonic
crystals {\it Phys. Rev. E} {\bf 64} 056604.

\bibitem{na} Akhmediev N,  Ankiewicz A and Taki M 2009 Waves that appear from nowhere and disappear without a trace {\it Phys. Lett. A}  {\bf 373} 675.

\bibitem{orw} Solli D R, Ropers C, Koonath P and Jalali B 2007 Optical rogue waves {\it Nature} {\bf 450} 1054.


\bibitem{orw2} Solli D R, Ropers C and Jalali B 2008 Active Control of RogueWaves for Stimulated Supercontinuum Generation {\it Phys. Rev. Lett.} {\bf 101} 233902.

\bibitem{orwt}  Vergeles S and Turitsyn S K 2011 Optical rogue waves in telecommunication data streams {\it Phys. Rev. A} {\bf 83} 061801(R).

\bibitem{srw} Dysthe K, Krogstad H E and M\"uller P 2008 Oceanic Rogue Waves {\it Annu. Rev. Fluid Mech.} {\bf 40} 287.

\bibitem{srw2} Osborne A R {\em Nonlinear Ocean Waves} (Academic Press, New York, 2009).

\bibitem{srw3} Kharif C and Pelinovsky E 2003 Physical mechanisms of the rogue wave phenomenon {\it Euro. J. Mech. B/Fluids} {\bf 22} 603-634.

\bibitem{srw4} Zakharov V E, Dyachenko A I and Prokofiev A O 2006 Freak waves as nonlinear stage of Stokes wave modulation instability {\it Euro. J. Mech. B/Fluids} {\bf 25} 677-692.

\bibitem{mrw} Bludov Yu V, Konotop V V and Akhmediev N 2009 Matter rogue waves {\it Phys. Rev. A}{ \bf 80} 033610.

\bibitem{mrw2} Yan Z Y, Konotop  V V and  Akhmediev N 2010 Three-dimensional rogue waves in nonstationary parabolic potentials {\it Phys. Rev. E}{ \bf 82} 033610.

\bibitem{frw} Yan Z Y 2010 Financial Rogue Waves {\it Commun. Theor. Phys.} {\bf 54} 947-949.[arXiv: 0911.4259]

\bibitem{frw2} Yan Z Y 2011 Vector financial rogue waves {\it Phys. Lett. A} {\bf 375} 4274-4279.

\bibitem{PS} Peregrine D H 1983 Water waves, nonlinear Schr\"odinger equations and their solutions
 {\it J. Austral. Math. Soc. Ser. B (Appl. Math.)} {\bf 25} 16.


\bibitem{nlsr} Akhmediev N, Ankiewicz A and Soto-Crespo J M 2009 Rogue waves and rational solutions of the nonlinear Schr\"odinger equation {\it Phys. Rev. E} {\bf 80} 026601.

\bibitem{hnls} Ankiewicz A, Devine N and Akhmediev N 2009 Are rogue waves robust against perturbations ? {\it Phys. Lett. A} {\bf 373} 3997-4000.

\bibitem{lilu} Yang G Y, Li L and Jia S T 2012 Peregrine rogue waves induced by the interaction between a continuous wave and a soliton {\it Phys. Rev. E}, {\bf 85}, 046608.

\bibitem{hr} Ankiewicz A, Soto-Crespo J M and Akhmediev N 2010 Rogue waves and rational solutions of the Hirota equation {\it Phys. Rev. E} {\bf 81} 046602.


\bibitem{drw} Ankiewicz A, Akhmediev N and  Soto-Crespo J M 2010 Discrete rogue waves of the Ablowitz-Ladik and Hirota equations {\it Phys. Rev. E} {\bf 82} 026602.

\bibitem{drw2} Akhmediev N and Ankiewicz A 2010 Modulation instability, Fermi-Pasta-Ulam recurrence, rogue waves, nonlinear phase shift, and exact solutions of the Ablowitz-Ladik equation {\it  Phys. Rev. E} {\bf 83} 046603.

\bibitem{drw3}  Yan Z Y and  Jiang D M 2012 Nonautonomous discrete rogue wave solutions and interactions in an
inhomogeneous lattice with varying coefficients {\it J. Math. Anal. Appl.} {\bf 395}, 542.


\bibitem{crw} Bludov Yu V, Konotop V V and Akhmediev N 2010 Vector rogue waves in binary mixtures
of Bose-Einstein condensates {\it Eur. Phys. J. Special Topics} {\bf 185} 169-180.

\bibitem{crw2} Baronio F {\it et al} 2012 Solutions of the vector nonlinear Schrodinger equations: evidence for deterministic rogue waves {\it Phys. Rev. Lett.} {\bf 109} 044102.

\bibitem{hrw} Dai C Q, Zhou G Q and Zhang J F 2012 Controllable optical rogue waves in the femtosecond regime {\it Phys. Rev. E} {\bf 85} 016603.

\bibitem{ghnls} Li H J {\it et al} 2008 High-order nonlinear Schr\"odinger equation and superluminal optical solitons
in room-temperature active-Raman-gain media {\it Phys. Rev. A} {\bf 78} 023822.

\bibitem{chow} Chow K W, Wong K K Y and Lama K 2008 Modulation instabilities in a system of four coupled, nonlinear Schr\"odinger equations {\it Phys. Lett. A} {\bf 372} 4596-4600.

\bibitem{lie} Bluman G W and Kumei S {\em Symmetries and differential equations} (Springer-Verlag, New York, 1989).

\bibitem{lie2} Bluman G W and Yan Z Y 2005 Nonclassical potential solutions of partial differential equations {\it Eur. J. Appl. Math.} {\bf 16} 239.

\bibitem{ps} Dysthe K B and Trulsen K 1999 Note on breather type solutions of the NLS asmodels for freak-waves {\it Phys. Scr.} {\bf T82} 48-52.


\bibitem{hnlsrw} Ankiewicz A,  Soto-Crespo J M and Akhmediev N 2010 Rogue waves and rational solutions of the Hirota equation {\it Phys. Rev. E} {\bf 81} 046602.


\bibitem{a68} Dai C Q, Wang Y Y, Tian Q and Zhang J F  2012 The management and containment of self-similar rogue waves in the inhomogeneous nonlinear Schrodinger equation {\it Ann. Phys.} {\bf 327} 512.

\bibitem{a69} Towers I and Malomed B A 2002 Stable (2+1)-dimensional solitons in a layered medium with sign-alternating Kerr nonlinearity
           {\it J. Opt. Soc. Am. B} {\bf 19} 537.




\end{references}
\end{document}